# Light-induced coherent interlayer transport in stripe-ordered $La_{1.6-x}Nd_{0.4}Sr_xCuO_4$

Morihiko Nishida[1], Kota Katsumi[1,2], Dongjoon Song[3], Hiroshi Eisaki[3], and Ryo Shimano[1,2]

[1] *Department of Physics, The University of Tokyo, Hongo, Tokyo 113-0033, Japan*
[2] *Cryogenic Research Center, The University of Tokyo, Yayoi, Tokyo 113-0032, Japan*
[3] *National Institute of Advanced Industrial Science and Technology, Tsukuba 305-8568, Japan*

**Abstract**

We have investigated the photoexcited transient responses of stripe-ordered phase in a cuprate superconductor, $La_{1.6-x}Nd_{0.4}Sr_xCuO_4$ ($x = 0.12$) using optical-pump terahertz (THz)-probe spectroscopy. Upon the near-infrared photoexcitation with the electric field polarized along the *c*-axis, a clear plasma edge appears in the THz reflection spectrum along the *c*-axis with its position nearly coinciding with the Josephson plasma resonance of similarly doped $La_{2-x}Sr_xCuO_4$ ($x = 0.125$) in the low-temperature superconducting phase. The appearance of light-induced plasma edge sustains up to the onset temperature of the charge-stripe order, indicating the inherent interplay between the light-induced phase and the charge-stripe order. The optical conductivity spectrum of the light-induced state is mostly reproduced by the Drude model with a scattering rate as small as a few meV, and its imaginary part does not exhibit $1/\omega$-divergence behavior in any temporal region after the photoexcitation. We discuss the possible origin of the observed coherent interlayer transport behavior as manifested by the narrow Drude response in the THz reflectivity along the *c*-axis.

**I. Introduction**

A common nature of high-$T_c$ cuprates is that the superconductivity emerges by doping carriers into parent compounds of antiferromagnetic Mott insulators. In the proximity of the superconductivity, various phases appear including the original antiferromagnetic phase, the pseudogap region, the charge- and spin-ordered phase, and the pair density wave order [1–5]. Elucidation of their interplay with the superconductivity has been one of the central issues in the study of high-$T_c$ superconducting cuprates. Recently, various experimental studies have revealed that the charge order is present in a wide range of high-$T_c$ cuprates and thus the relevance of the charge order to the superconductivity and other proximity phases has been intensively investigated [5–7]. One of the most concrete forms of the charge order is perhaps the stripe order [8,9], in which both charges and spins are spontaneously segregated in a form of stripes in the $CuO_2$ plane, as found in 214 families of La-based cuprates including $La_{2-x}Ba_xCuO_4$ (LBCO), $La_{2-x-y}Eu_ySr_xCuO_4$ (LESCO), and $La_{2-x-y}Nd_ySr_xCuO_4$ [9,10]. Near the hole doping level of $x = 1/8$, the static charge- and spin-stripe orders (CO and SO, respectively) are stabilized by pinning to the low-temperature tetragonal (LTT) lattice deformation [8–12], associated with a profound suppression of the 3D-superconducting critical temperature $T_c$ (Fig. 1(a)). This phenomenon, dubbed 1/8-anomaly, primarily indicates the strong competition between the stripe orders and bulk superconductivity, whereas several experimental results such as neutron scattering [13] and resistivity measurements [14,15] have revealed the coexistence of superconductivity and stripe orders. The interplay between the charge-stripe order and superconductivity has also been investigated by resonant x-ray scattering [16–18]. These experiments indicate the intimate connection between the stripe orders and superconductivity.

Recent developments in ultrafast pump-probe spectroscopy techniques have provided further insights into the relationship between the stripe orders and superconductivity. Remarkably, the emergence of plasma edge in terahertz (THz) reflectivity spectrum was observed in 1/8-doped LESCO above $T_c$ up to the onset temperature of CO ($T_{CO}$) after the mid-infrared pulse excitation and interpreted as light-induced superconducting Josephson plasma edge [19,20]. Similar behavior was identified in the stripe phase of the LBCO system by mid-infrared and also by visible wavelength optical pumping [21,22]. These phenomena were interpreted as a recovery of $c$-axis Josephson coupling caused by the photo-induced destruction of the stripe order, indicating the development of 2D-superconducting pairings in the charge-ordered state in equilibrium. The presence of equilibrium in-plane superconductivity above $T_c$ was also indicated from the observation of THz third-harmonic generation along the $c$-axis [23,24].

In this paper, to have a deeper insight into the intertwining of the charge- and spin-stripe order and the superconductivity, we applied the ultrafast spectroscopy technique to another stripe-ordered system of La-214 family, $La_{1.6-x}Nd_{0.4}Sr_xCuO_4$ (LNSCO), where the static charge and spin stripes are stabilized by $Nd^{3+}$ substitution [10,13,25–28]. X-ray scattering study has revealed a shorter correlation length in LNSCO than in LBCO [29]. The study of how such a short-range character of charge stripes influences the light-induced superconducting-like state is imperative for the understanding of the interplay between those competing or coexisting orders.

## II. Method

We used a La$_{1.6-x}$Nd$_{0.4}$Sr$_x$CuO$_4$ single crystal with the doping level of $x = 0.12$ grown by the floating-zone method, where $T_{CO} = 67$ K and $T_{SO} = 55$ K [25]. $T_c$ was determined as 3.2 K from the onset temperature of diamagnetism. All the pump-probe experiments were performed above $T_c$. The mirror-polished *ac* surface of the sample ($\phi$5 mm) was mounted on a copper sample holder with a tapered hole ($\phi$3 mm). In Fig. 1(b) we show a schematic of near-infrared optical-pump THz-probe spectroscopy. As a light source, we used a Ti:Sapphire-based regenerative amplifier with a pulse energy of 7 mJ, a repetition rate of 1 kHz, a pulse duration of 100 fs, and a center wavelength of 800 nm (1.55 eV in photon energy). The output of the laser was divided into three beams: each for the optical pump, probe THz generation, and the gate pulse for the THz time-domain spectroscopy, respectively. The optical pump beam has a Gaussian profile with a $1/e^2$ diameter of 6 mm, which ensures spatially uniform excitation on the probed region. The THz probe pulse was generated by the optical rectification in a large-aperture ZnTe crystal. Both the optical pump and THz probe pulse were linearly polarized along the *c*-axis of LNSCO. The reflected THz pulse was detected by the electro-optic sampling in a ZnTe crystal with the gate pulse. To obtain the pump-induced reflection spectrum change, the delay time between the pump pulse and the gate pulse ($t_{pp}$) was fixed and the probe THz pulse was swept relative to the gate pulse, and then Fourier-transformed.

## III. Results

Figure 2(a) shows the light-induced *c*-axis reflectivity change for each pump-probe delay $t_{pp}$ with a pump fluence of 1 mJ/cm$^2$ at 10 K. While the reflection spectrum is featureless in equilibrium in the measured photon energy range as shown in the inset, a sharp reflectivity edge emerges, and then redshifts within a few picoseconds after the photoexcitation. To capture the dynamics of the photo-induced state in a longer time scale, we plotted the reflected THz electric field change $\Delta E/E$ as a function of $t_{pp}$ in Fig. 2(b). After the photoexcitation, $\Delta E/E$ first rises with a slight delay (~ 1 ps) relative to the pump pulse, followed by exponential decaying behavior with two characteristic time scales: 2 ps for the fast one and much longer than 100 ps for the slow one.

To take into account the effect of penetration depth mismatch between the optical pump and THz probe pulses, we adopted a multi-layer model [30,31] and extracted the optical response functions at the sample surface, where the pump-induced reflectivity change is the largest. In this model, the induced refractive index change is assumed to decay exponentially along the sample's depth direction, as given by $n(z) = (n_{surf} - n_{eq}) \exp(-z/d) + n_{eq}$. Here, $z$ is the depth from the surface, $d = 0.31$ μm is the penetration depth for the optical pump pulse determined by ellipsometry measurements, $n_{surf}$ is the refractive index at the surface region under the photoexcitation, and $n_{eq} = 4.7$ is the refractive index in equilibrium without the pump which was extracted from the reflection spectrum shown in the inset of Fig. 2(a), by approximating that it is constant in the measured photon energy range. We note here that the spectral analysis using the above exponential decay model can result in artificial spectral structures in particular below $T_c$ unless the effect of heating is correctly taken into account [32], but such an artifact is not significant in the present LNSCO system above $T_c$ because the equilibrium reflection spectrum is featureless.

Figure 3(a) presents surface reflection spectra obtained from $n_{surf}$ using Fresnel's formula for the reflectivity, $R = |(1 - n_{surf}) / (1 + n_{surf})|^2$. A sharp edge structure arises with its position almost identical to that of the raw transient reflection spectrum shown in Fig. 2(a). The induced surface reflectivity approaches unity toward the low energy limit, which is in accordance with the behavior expected for the plasma edge.

Corresponding to the edge structure observed in the reflection spectra, a sharp peak is identified in the loss function spectra defined by $-\text{Im}[1/\varepsilon(\omega)]$ where $\varepsilon(\omega) = n_{surf}^2(\omega)$ is the dielectric function, as represented in Fig. 3(b), which corresponds to the longitudinal plasma resonance. This peak also shows a redshift after the photoexcitation and moves away from the measured spectral range after 20 ps. As a reference, the $c$-axis reflectivity and the loss function spectrum of $La_{2-x}Sr_xCuO_4$ (LSCO) with $x = 0.125$ sample in its equilibrium superconducting state at 5 K are shown by the black dashed lines.

From the complex transient reflection spectrum obtained by the THz time-domain spectroscopy, the photo-induced changes of the real and imaginary part of the optical conductivity, $\Delta\sigma_1(\omega)$ and $\Delta\sigma_2(\omega)$, are extracted without using the Kramers-Kronig analysis as shown in Figs. 3(c) and 3(d), respectively, at the indicated pump-probe delay. The spectral features of the induced optical conductivity, $\Delta\sigma_1(\omega)$ and $\Delta\sigma_2(\omega)$, for later delays at $t_{pp} = 5.8$ ps and 20 ps are reasonably well described by the Drude model, $\Delta\sigma_1(\omega) + i\Delta\sigma_2(\omega) = i\varepsilon_0\omega_p^2/(\omega + i\gamma)$, as represented by the bold lines in Figs. 3(c) and 3(d), where $\varepsilon_0$ is the vacuum permittivity, $\omega_p$ is the plasma frequency, and $\gamma$ is the scattering rate. A deviation from the Drude model is discerned for earlier delays at $t_{pp} = 1.0\text{-}2.6$ ps, but the $1/\omega$-divergent behavior is not observed in the $\Delta\sigma_2(\omega)$ spectra at any $t_{pp}$ unlike the case of LBCO and LESCO [19–22]. As a possible origin for the peak structure that appeared in $\Delta\sigma_1(\omega)$, one may infer the spatial localization of the charge carriers which should transfer the low-energy spectral weight to the high-energy side [33]. This may be related to the short-ranged character of charge stripes in LNSCO [29]. We will return to this point later.

One may also consider the artifact of the ultrafast pump-probe signal that appears when the signal decays faster than the period of probe THz field [34]. To examine this possibility, we performed the optical-pump THz-probe spectroscopy with a longer pump pulse duration ranging from 2.4 ps to 6.8 ps using a grating pair (see Appendix A for the experimental details). As displayed in Figs. 4(a) and 4(b), the loss function peak corresponding to the longitudinal plasma resonance evolves after the pump pulse irradiation accompanied by a slight delay (~ 1 ps) as in the case of short-pulse (100-fs) excitation, which may reflect the buildup timescale of the plasmon, or the melting dynamics of charge stripes [17] as we discuss later. Importantly, the loss function peak reaches around 4 meV at the largest and then shows a redshift, as also observed in the short-pulse excitation experiments. These results indicate that the emergence of the plasmon peak reflects the intrinsic nature of the light-induced nonequilibrium state, and its lifetime can be prolonged by using a longer pump pulse. The prolongation of the light-induced state is also reported in the $YBa_2Cu_3O_{6+x}$ system recently [35], although the microscopic mechanism for realizing the light-induced state may differ from that of the stripe-ordered system.

Next, we show the temperature dependences of the light-induced reflectivity change and the loss function spectrum at $t_{pp}$ = 1.0 ps in Figs. 5(a) and 5(b), respectively. With increasing temperature, the plasma edge and the correspondent loss function peak show a redshift and disappear above $T_{CO}$. Also shown in Figs. 5(c) and 5(d) are $\Delta\sigma_1(\omega)$ and $\Delta\sigma_2(\omega)$ spectra, respectively for each temperature at $t_{pp}$ = 1.0 ps. $\Delta\sigma$ gradually decreases as temperature increases but sustains up to $T_{CO}$ with its spectral shape nearly unchanged. The dynamics of the reflected THz electric field change $\Delta E/E(t_{pp})$ is displayed in Fig. 5(e) for each temperature. The data were fitted by the following function:

$$f(t_{pp}) = \frac{1 + \text{erf}[(t_{pp} - t_0)/\sigma]}{2} \left[ A_1 \exp\left(-\frac{t_{pp} - t_0}{\tau}\right) + A_2 \right], \quad (1)$$

where the fast-decaying component is described by the amplitude $A_1$ and the lifetime $\tau$, and the long-lived component is approximated by a constant term $A_2$. The extracted temperature dependence of the parameters $A_1$ and $A_2$ are summarized in Fig. 5(f). We can clearly see that the fast-decaying component $A_1$ develops below $T_{CO}$, while $A_2$ develops at a lower temperature.

**IV. Discussion**

Now we discuss the possible origin of the observed photo-induced plasma edge and the coherent interlayer transport behavior as indicated by the narrow Drude response in the THz reflectivity along the *c*-axis. Because the reflection spectrum in equilibrium is nearly featureless in the THz frequency range due to the insulating character along the *c*-axis in LNSCO above $T_c$, the emergence of such a distinct plasma edge structure by the photoexcitation cannot be attributed to the heating effect. In the present case of the 1.55 eV pump polarized along the *c*-axis, the initial process of the photoexcitation should be primarily the charge redistribution, most likely the transfer of holes from $CuO_2$ planes to apical oxygens as reported in Ref. [36]. However, the photo-doped carriers themselves cannot be directly attributed to the origin of the observed Drude weight because the observed weight decreases with increasing temperature and disappears above $T_{CO}$.

Combined with the previous observation in LESCO and LBCO systems, the light-induced reflectivity edge is interpreted as a universal character of the charge-stripe phase in cuprates. As suggested from recent time-resolved resonant x-ray scattering measurements in the LBCO system in which the CO melting is shown to occur within 1 ps after the photoexcitation with a recovery time of about 5 ps [17], it is likely that the melting of CO also occurs in the present case of LNSCO. In fact, the pump fluence of 1 mJ/cm$^2$ (corresponding to $1 \times 10^{-2}$ photon/Cu) in our experiments is higher than the previously investigated pump fluence region for the CO melting to occur [17].

Unlike the case of LBCO and LESCO [19–22], however, the $1/\omega$-divergence is not observed in $\Delta\sigma_2$ in the present case of LNSCO, and instead, a narrow Drude response appears in the earlier time (~ 1 ps) after the photoexcitation. One may consider that the observed narrow Drude response is ascribed to quasiparticles that appear when the CO is destroyed. In fact, the appearance of the narrow Drude response itself has been reported as an in-plane response in the normal state of the underdoped $HgBa_2CuO_{4+\delta}$ with exhibiting the Fermi liquid behavior [37]. Therefore, one can attribute in principle the origin of the light-induced Drude weight along the *c*-axis to quasiparticles in the normal state. Although we do not rule out this possibility, it is highly nontrivial that such coherent quasiparticles along the *c*-axis with a scattering rate as small as 3 meV appear below 70 K associated with the destruction of the CO.

Taking into account that the static CO is a competing order with the superconductivity, it is still tempting to relate the observed coherent *c*-axis interlayer transport behavior to the *c*-axis pair tunneling that is activated by quenching the CO. As we have seen in Fig. 3, the coincidence of the energy of the light-induced plasma edge in LNSCO and that of LSCO in the equilibrium superconducting phase in similar doping ($x = 0.125$) is indicative. Along with this picture, the recent theoretical investigation using variational Monte-Carlo calculation has shown the transient appearance of the superconductivity after the photoexcitation with suppressing the charge order which dominates the superconductivity in equilibrium as a consequence of a subtle free energy competition between the CO and the superconductivity [38]. In light of this picture, we then infer that the unusually narrow Drude response that appeared after the photoexcitation is still related to the superconductivity even though the $1/\omega$-like divergence is not observed in $\Delta\sigma_2$.

Suggestively, a recent theoretical study has shown that the *d*-wave superconducting order is enhanced by an optical pump while its superconducting correlation is limited to a short-range one, associated with the suppression of antiferromagnetic spin order [39]. Such short-range nonequilibrium superconductivity may explain the suppression of $1/\omega$-divergence in $\Delta\sigma_2(\omega)$ in the low energy (frequency) range while maintaining the Josephson plasma edge structure in the THz frequencies [40]. It is also noteworthy that the residual Drude weight of in-plane THz response with a relatively small scattering rate of ~ 3 meV has been observed in thin-film samples of overdoped LSCO system below $T_c$ in equilibrium and attributed to phase fluctuating Cooper pairs [41]. Although it is not clear that the same argument can be applied to the *c*-axis response of LNSCO, the observed narrow Drude response in the photo-induced state may indicate the fluctuating superconductivity [42]. This behavior in LNSCO distinct from LBCO may share a common origin with the short correlation length of charge stripes observed in LNSCO [29].

Finally, we comment on the possible origin of the long-lived component described by the $A_2$ term in Eq. (1). As shown in Fig. 3(a), even in such a longer delay limit, the edge-like behavior is also discerned at the low energy region to which the high energy plasma edge structure asymptotically approaches. Presumably, such a long-lived signal is attributed to the heating effect caused by photoexcitation. Using the heat capacity of LNSCO [43], the temperature increase by the pump pulse with the fluence of 1 mJ/cm$^2$ is estimated to be about 60 K. Therefore, the charge stripes are considered to be melt thermally in a longer time delay, which may explain the increase of the conductivity in such a long-lived component. This long-lived behavior also deserves further investigation to clarify the slow dynamics of charge-stripe order.

## V. Conclusion

We have investigated the photoexcitation effect of the stripe-ordered phase in La$_{1.6-x}$Nd$_{0.4}$Sr$_x$CuO$_4$ ($x = 0.12$) above $T_c$ by optical pump-THz probe spectroscopy along the $c$-axis. The transient reflection spectrum shows a clear plasma edge in the THz frequency range and its maximum position nearly coincides with that of La$_{2-x}$Sr$_x$CuO$_4$ ($x = 0.125$) in its equilibrium superconducting phase. This photo-induced plasma edge was identified only below $T_{CO}$, indicating that the suppression of the charge stripes is crucial for the emergence of the photo-induced plasma edge.

The extracted transient complex optical conductivity exhibits a Drude-like spectrum with a small scattering rate of $\gamma \sim 3$ meV. This coherent charge carrier response along the $c$-axis is discussed in view of fluctuating superconductivity with a short-range correlation length. It is indicative that such a light-induced plasma edge above $T_c$ was not observed in the LSCO system [32]. This difference may suggest that the charge instability that causes the charge stripes in LNSCO is intimately related to the enhancement of the pairing instability which manifests itself in a form of coherent $c$-axis interlayer transport when the static charge-stripe order is quenched off by the photoexcitation.

**Acknowledgement**

We acknowledge N. Yoshikawa, H. Niwa, T. Tomiyasu, R. Suda, K. Konishi, and T. Fujii for fruitful discussions and experimental assistance. This work was partly supported by JSPS KAKENHI Grants No. 18H05324 and No. 15H02102 and JST CREST Grant No. JPMJCR19T3.

## Appendix A: Long-Pulse Excitation with a Chirped Pulse

To confirm that the observed plasma edge and the loss function peak are not sensitive to the pump pulse duration, we performed optical-pump THz-probe spectroscopy with a prolonged pump pulse. For this purpose, we used a Ti:Sapphire-based regenerative amplifier with a pulse energy of 2 mJ and pulse duration of 35 fs. As schematically illustrated in Fig. 6, the beam output was divided into optical pump, THz generation, and gate pulse. The optical pump pulse was down-chirped by a grating pair. The pulse width is tuned between 2.4-6.8 ps by changing the grating separation. The pump fluence is 3 and 4 mJ/cm$^2$ for 2.4 and 6.8 ps pumping, respectively. These fluences were determined as the value at which the pump-induced signal begins to saturate, by the same procedure as the short-pulse excitation. The other parameters are the same as the 100−fs pumping setup.

**Appendix B: Pump-Fluence Dependence**

To study further the behavior of the photo-induced state, we have measured the pump-fluence dependence of the reflectivity change and the loss function spectra as represented in Figs. 7(a) and 7(b), respectively, at $t_{pp}$ = 1.0 ps. With increasing the pump fluence, the plasma edge in the reflection spectrum blue-shifts and asymptotically approaches the maximum value of ~ 4 meV. Concomitantly the loss function peak shows a blue shift with increasing pump fluence, whose energy is plotted in the inset of Fig. 7(b) against the pump fluence. These behaviors are consistent with the long-pulse excitation experiments described in the main text, where the increase of the loss function peak energy was observed following the rising of the pump pulse envelope.

The fluence dependences of the photo-induced complex optical conductivity at $t_{pp}$ = 1.0 ps are also displayed in Figs. 7(c) and 7(d). The spectral shapes of $\Delta\sigma_1(\omega)$ and $\Delta\sigma_2(\omega)$ are almost independent of the pump fluence and show a saturation behavior above 0.5 mJ/cm$^2$.

In our analysis of the surface layer model, the saturation effect of the pump-induced refractive index change discussed in Ref. [44] is not taken into account, because the loss function peak energy grows linearly up to ~ 3 meV with the pump fluence as shown in the inset of Fig.7(b). Above the fluence of 0.5 mJ/cm$^2$, we can discern the saturation behavior in the extracted optical responses. If the saturation of the refractive index change is taken into account, the effective pump penetration depth should become longer (but yet sufficiently shorter than the THz probe penetration depth). This correction of the penetration depth should cause the more pronounced saturation of the plasma edge energy in such a high fluence region, and thus we can conclude that the asymptotic value of the plasma resonance should rest between 3 and 4 meV.

**References**

[1]  B. Keimer, S. A. Kivelson, M. R. Norman, S. Uchida, and J. Zaanen, *From Quantum Matter to High-Temperature Superconductivity in Copper Oxides*, Nature **518**, 179 (2015).

[2]  T. Yoshida, M. Hashimoto, I. M. VIshik, Z. X. Shen, and A. Fujimori, *Pseudogap, Superconducting Gap, and Fermi Arc in High-T$_c$ Cuprates Revealed*

*by Angle-Resolved Photoemission Spectroscopy*, J. Phys. Soc. Japan **81**, 011006 (2012).

[3] M. Fujita, H. Hiraka, M. Matsuda, M. Matsuura, J. M. Tranquada, S. Wakimoto, G. Xu, and K. Yamada, *Progress in Neutron Scattering Studies of Spin Excitations in High-$T_c$ Cuprates*, J. Phys. Soc. Japan **81**, 011007 (2012).

[4] D. F. Agterberg, J. C. S. Davis, S. D. Edkins, E. Fradkin, D. J. Van Harlingen, S. A. Kivelson, et al., *The Physics of Pair-Density Waves: Cuprate Superconductors and Beyond*, Annu. Rev. Condens. Matter Phys. **11**, 231 (2020).

[5] S. Uchida, *Ubiquitous Charge Order Correlations in High-Temperature Superconducting Cuprates*, J. Phys. Soc. Japan **90**, 111001 (2021).

[6] R. Arpaia and G. Ghiringhelli, *Charge Order at High Temperature in Cuprate Superconductors*, J. Phys. Soc. Japan **90**, 111005 (2021).

[7] R. Comin and A. Damascelli, *Resonant X-Ray Scattering Studies of Charge Order in Cuprates*, Annu. Rev. Condens. Matter Phys. **7**, 369 (2016).

[8] J. Zaanen and O. Gunnarsson, *Charged Magnetic Domain Lines and the Magnetism of High-$T_c$ Oxides*, Phys. Rev. B **40**, 7391 (1989).

[9] J. M. Tranquada, *Cuprate Superconductors as Viewed through a Striped Lens*, Adv. Phys. **69**, 437 (2020).

[10] J. M. Tranquada, B. J. Sternlieb, J. D. Axe, Y. Nakamura, and S. Uchida, *Evidence for Stripe Correlations of Spins and Holes in Copper Oxide Superconductors*, Nature **375**, 561 (1995).

[11] J. Fink, E. Schierle, E. Weschke, J. Geck, D. Hawthorn, V. Soltwisch, et al., *Charge Ordering in $La_{1.8-x}Eu_{0.2}Sr_xCuO_4$ Studied by Resonant Soft x-Ray Diffraction*, Phys. Rev. B **79**, 100502(R) (2009).

[12] T. J. Boyle, M. Walker, A. Ruiz, E. Schierle, Z. Zhao, F. Boschini, et al., *Large Response of Charge Stripes to Uniaxial Stress in $La_{1.475}Nd_{0.4}Sr_{0.125}CuO_4$*, Phys. Rev. Res. **3**, L022004 (2021).

[13] J. M. Tranquada, J. D. Axe, N. Ichikawa, A. R. Moodenbaugh, Y. Nakamura, and S. Uchida, *Coexistence of, and Competition between, Superconductivity and Charge-Stripe Order in $La_{1.6-x}Nd_{0.4}Sr_xCuO_4$*, Phys. Rev. Lett. **78**, 338 (1997).

[14] Q. Li, M. Hücker, G. D. Gu, A. M. Tsvelik, and J. M. Tranquada, *Two-Dimensional Superconducting Fluctuations in Stripe-Ordered $La_{1.875}Ba_{0.125}CuO_4$*, Phys. Rev. Lett. **99**, 067001 (2007).

[15] J. M. Tranquada, G. D. Gu, M. Hücker, Q. Jie, H. J. Kang, R. Klingeler, et al., *Evidence for Unusual Superconducting Correlations Coexisting with Stripe Order in $La_{1.875}Ba_{0.125}CuO_4$*, Phys. Rev. B **78**, 174529 (2008).

[16] M. Först, R. I. Tobey, H. Bromberger, S. B. Wilkins, V. Khanna, A. D. Caviglia, et al., *Melting of Charge Stripes in Vibrationally Driven $La_{1.875}Ba_{0.125}CuO_4$: Assessing the Respective Roles of Electronic and Lattice Order in Frustrated Superconductors*, Phys. Rev. Lett. **112**, 157002 (2014).

[17] M. Mitrano, S. Lee, A. A. Husain, L. Delacretaz, M. Zhu, G. De La Peña Munoz, et al., *Ultrafast Time-Resolved x-Ray Scattering Reveals Diffusive Charge Order Dynamics in $La_{2-x}Ba_xCuO_4$*, Sci. Adv. **5**, eaax3346 (2019).

[18] M. Bluschke, N. K. Gupta, H. Jang, A. A. Husain, B. Lee, M. Na, et al., *Nematicity Dynamics in the Charge-Density-Wave Phase of a Cuprate Superconductor*, arXiv:2209.11528.

[19] D. Fausti, R. I. Tobey, N. Dean, S. Kaiser, A. Dienst, M. C. Hoffmann, et al., *Light-Induced Superconductivity in a Stripe-Ordered Cuprate*, Science **331**, 189 (2011).

[20] C. R. Hunt, D. Nicoletti, S. Kaiser, T. Takayama, H. Takagi, and A. Cavalleri, *Two Distinct Kinetic Regimes for the Relaxation of Light-Induced Superconductivity in $La_{1.675}Eu_{0.2}Sr_{0.125}CuO_4$*, Phys. Rev. B **91**, 020505(R) (2015).

[21] D. Nicoletti, E. Casandruc, Y. Laplace, V. Khanna, C. R. Hunt, S. Kaiser, et al., *Optically Induced Superconductivity in Striped $La_{2-x}Ba_xCuO_4$ by Polarization-Selective Excitation in the near Infrared*, Phys. Rev. B **90**, 100503(R) (2014).

[22] E. Casandruc, D. Nicoletti, S. Rajasekaran, Y. Laplace, V. Khanna, G. D. Gu, J. P. Hill, and A. Cavalleri, *Wavelength-Dependent Optical Enhancement of Superconducting Interlayer Coupling in $La_{1.885}Ba_{0.115}CuO_4$*, Phys. Rev. B **91**, 174502 (2015).

[23] S. Rajasekaran, J. Okamoto, L. Mathey, M. Fechner, V. Thampy, G. D. Gu, and A. Cavalleri, *Probing Optically Silent Superfluid Stripes in Cuprates*, Science **359**, 575 (2018).

[24] D. Fu, D. Nicoletti, M. Fechner, M. Buzzi, G. D. Gu, and A. Cavalleri, *Terahertz Phase Slips in Striped $La_{2-x}Ba_xCuO_4$*, Phys. Rev. B **105**, L020502 (2022).

[25] J. M. Tranquada, J. D. Axe, N. Ichikawa, Y. Nakamura, S. Uchida, and B. Nachumi, *Neutron-Scattering Study of Stripe-Phase Order of Holes and Spins in $La_{1.48}Nd_{0.4}Sr_{0.12}CuO_4$*, Phys. Rev. B **54**, 7489 (1996).

[26] M. v. Zimmermann, A. Vigliante, T. Niemöller, N. Ichikawa, T. Frello, J. Madsen, et al., *Hard–X-Ray Diffraction Study of Charge Stripe Order in $La_{1.48}Nd_{0.4}Sr_{0.12}CuO_4$*, Europhys. Lett. **41**, 629 (1998).

[27] T. Niemöller, N. Ichikawa, T. Frello, H. Hünnefeld, N. H. Andersen, S. Uchida, J. R. Schneider, and J. M. Tranquada, *Charge Stripes Seen with X-Rays in*

$La_{1.45}Nd_{0.4}Sr_{0.15}CuO_4$, Eur. Phys. J. B **12**, 509 (1999).

[28] N. Ichikawa, S. Uchida, J. M. Tranquada, T. Niemöller, P. M. Gehring, S. H. Lee, and J. R. Schneider, *Local Magnetic Order vs Superconductivity in a Layered Cuprate*, Phys. Rev. Lett. **85**, 1738 (2000).

[29] S. B. Wilkins, M. P. M. Dean, J. Fink, M. Hücker, J. Geck, V. Soltwisch, et al., *Comparison of Stripe Modulations in $La_{1.875}Ba_{0.125}CuO_4$ and $La_{1.48}Nd_{0.4}Sr_{0.12}CuO_4$*, Phys. Rev. B **84**, 195101 (2011).

[30] A. Pashkin, M. Porer, M. Beyer, K. W. Kim, A. Dubroka, C. Bernhard, et al., *Femtosecond Response of Quasiparticles and Phonons in Superconducting $YBa_2Cu_3O_{7-\delta}$ Studied by Wideband Terahertz Spectroscopy*, Phys. Rev. Lett. **105**, 067001 (2010).

[31] S. Kaiser, C. R. Hunt, D. Nicoletti, W. Hu, I. Gierz, H. Y. Liu, et al., *Optically Induced Coherent Transport Far above $T_c$ in Underdoped $YBa_2Cu_3O_{6+\delta}$*, Phys. Rev. B **89**, 184516 (2014).

[32] H. Niwa, N. Yoshikawa, K. Tomari, R. Matsunaga, D. Song, H. Eisaki, and R. Shimano, *Light-Induced Nonequilibrium Response of the Superconducting Cuprate $La_{2-x}Sr_xCuO_4$*, Phys. Rev. B **100**, 104507 (2019).

[33] T. Timusk, D. N. Basov, C. C. Homes, A. V. Puchkov, and M. Reedyk, *Gap States in HTSC by Infrared Spectroscopy*, J. Supercond. **8**, 437 (1995).

[34] J. Orenstein and J. S. Dodge, *Terahertz Time-Domain Spectroscopy of Transient Metallic and Superconducting States*, Phys. Rev. B **92**, 134507 (2015).

[35] A. Ribak, M. Buzzi, D. Nicoletti, R. Singla, Y. Liu, S. Nakata, B. Keimer, and A. Cavalleri, *Two-Fluid Dynamics in Driven $YBa_2Cu_3O_{6.48}$*, arXiv:2210.08539.

[36] T. Tang, Y. Wang, B. Moritz, and T. P. Devereaux, *Orbitally Selective Resonant Photodoping to Enhance Superconductivity*, Phys. Rev. B **104**, 174516 (2021).

[37] S. I. Mirzaei, D. Stricker, J. N. Hancock, C. Berthod, A. Georges, E. Van Heumen, et al., *Spectroscopic Evidence for Fermi Liquid-like Energy and Temperature Dependence of the Relaxation Rate in the Pseudogap Phase of the Cuprates*, Proc. Natl. Acad. Sci. U. S. A. **110**, 5774 (2013).

[38] K. Ido, T. Ohgoe, and M. Imada, *Correlation-Induced Superconductivity Dynamically Stabilized and Enhanced by Laser Irradiation*, Sci. Adv. **3**, e1700718 (2017).

[39] Y. Wang, T. Shi, and C. C. Chen, *Fluctuating Nature of Light-Enhanced d-Wave Superconductivity: A Time-Dependent Variational Non-Gaussian Exact Diagonalization Study*, Phys. Rev. X **11**, 041028 (2021).

[40] Y. Lemonik and A. Mitra, *Quench Dynamics of Superconducting Fluctuations*

and *Optical Conductivity in a Disordered System*, Phys. Rev. B **98**, 214514 (2018).

[41] F. Mahmood, X. He, I. Božović, and N. P. Armitage, *Locating the Missing Superconducting Electrons in the Overdoped Cuprates $La_{2-x}Sr_xCuO_4$*, Phys. Rev. Lett. **122**, 027003 (2019).

[42] We should also note here that an alternative interpretation has been made to explain the in-plane narrow Drude response observed in the overdoped region of LSCO in terms of residual quasiparticles response originated from disorder effect as reported in N. R. Lee-Hone, V. Mishra, D. M. Broun, and P. J. Hirschfeld, *Optical Conductivity of Overdoped Cuprate Superconductors: Application to $La_{2-x}Sr_xCuO_4$*, Phys. Rev. B **98**, 054506 (2018), and the interpretation of the in-plane narrow Drude response in particular for the overdoped region of LSCO remains under debate. However, this quasiparticle picture fails to explain the appearance of plasma edge in the photo-induced LNSCO system that eventually coincides with that in a similarly doped LSCO system in the equilibrium superconducting state.

[43] J. Takeda, T. Inukai, and M. Sato, *Electronic Specific Heat of $(La,Nd)_{2-x}Sr_xCu_{1-y}Zn_yO_4$ up to about 300 K*, J. Phys. Chem. Solids **62**, 181 (2001).

[44] J. S. Dodge, L. Lopez, and D. G. Sahota, *Photoinduced Superconductivity Reconsidered: The Role of Photoconductivity Profile Distortion*, arXiv:2210.01114.

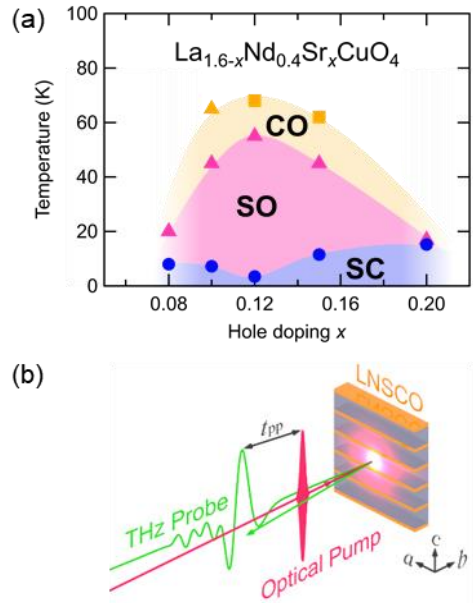

FIG. 1. (a) Phase diagram of $La_{1.6-x}Nd_{0.4}Sr_xCuO_4$ (LNSCO). CO: Charge-stripe order observed by x-ray (orange squares, [26,27]) and neutron (orange triangle, [28]) diffraction, SO: Spin-stripe order by neutron scattering (magenta triangles, [13,25,28]), SC: Superconducting state by magnetic susceptibility (blue circles, [13,28]). (b) Schematic of the optical pump-THz probe measurement along the $c$-axis of LNSCO.

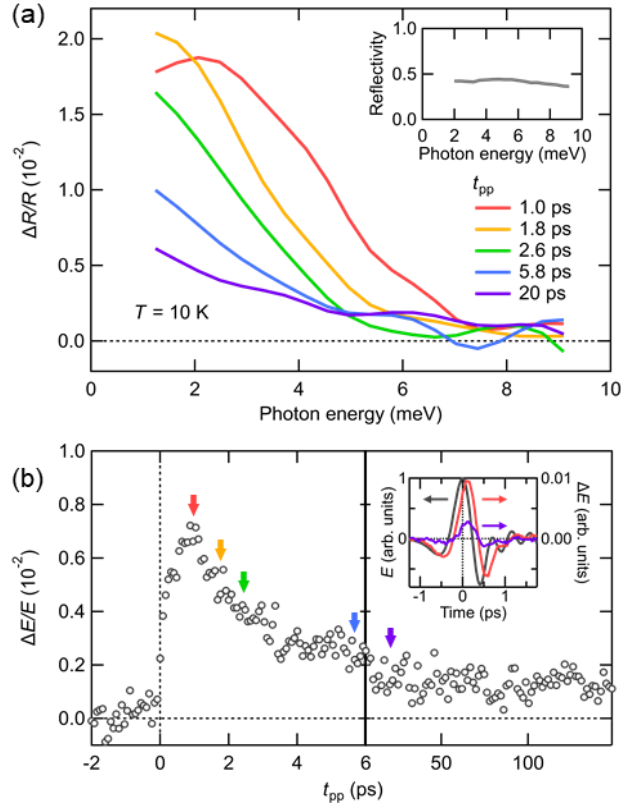

FIG. 2. Dynamics of reflectivity change after the photoexcitation at 10 K. (a) Transient reflectivity change spectra. Inset shows equilibrium reflectivity spectrum at 10 K. (b) Dynamics of THz electric field change $\Delta E/E$. Colored arrows indicate the pump-probe delays corresponding to the traces in (a). Inset: reflected THz waveform without pump, $E$ (gray line), and pump-induced electric field change $\Delta E$ for $t_{pp} = 1.0$ ps (red line) and 20 ps (purple line). $\Delta E/E$ in the main figure was measured at the peak position of the probe THz $E$-field.

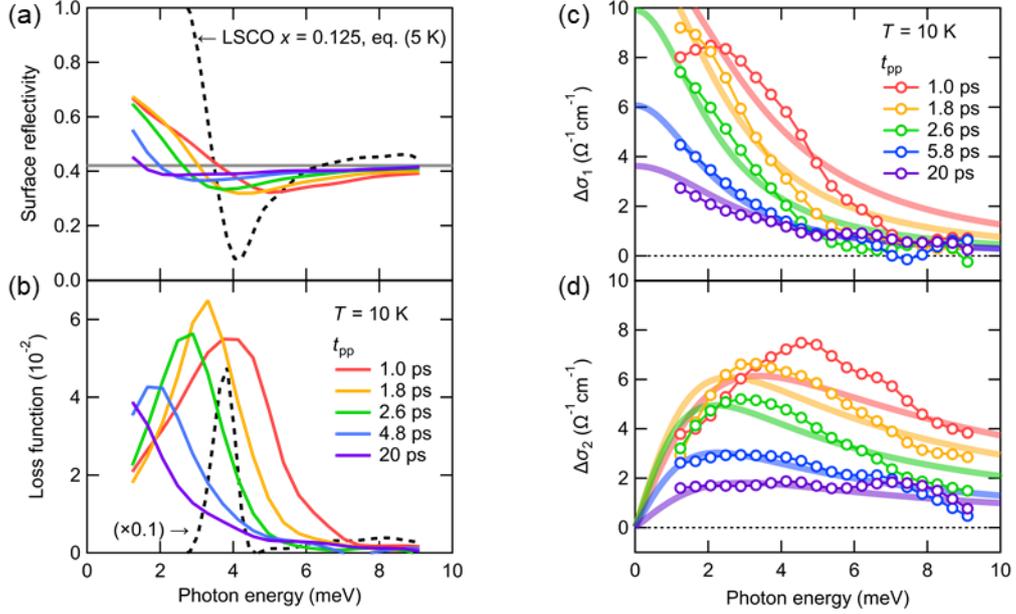

FIG. 3. Optical spectra at the sample surface extracted by the multi-layer model. (a) Surface reflectivity, defined as the reflectivity of the boundary between air and photoexcited sample surface. Gray horizontal line represents the equilibrium reflectivity calculated from $n_{eq} = 4.7$. (b) Loss function spectra, $-\text{Im}[1/\varepsilon(\omega)]$. The peak corresponds to the longitudinal plasma resonance. The black dashed lines in (a) and (b) indicate the equilibrium reflectivity and loss function spectrum of superconducting LSCO with a doping level of $x = 0.125$ at 5 K. (c, d) Real and imaginary part of the conductivity change. Bold lines are fitting results by the Drude model.

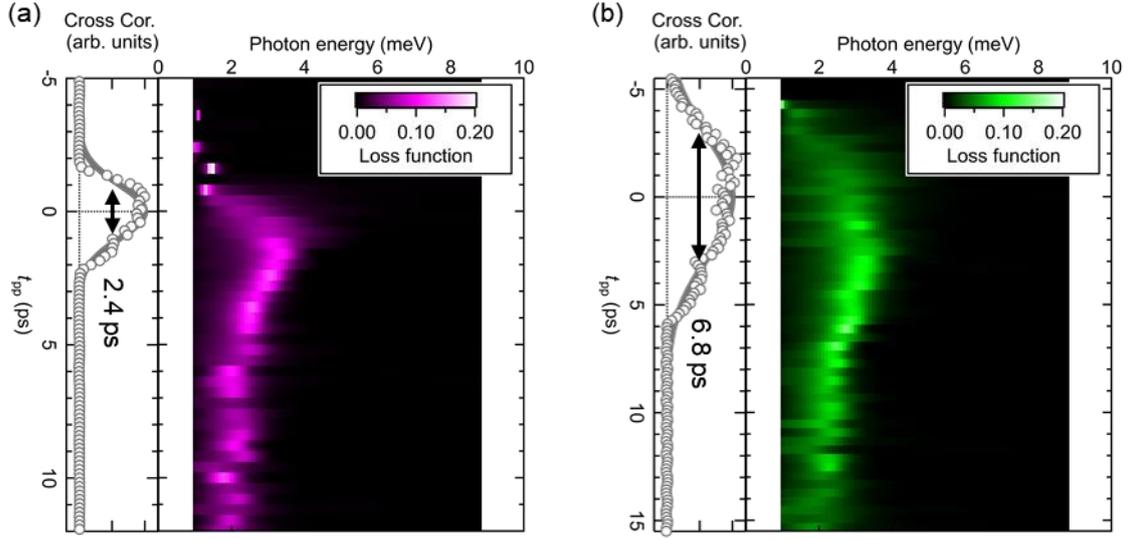

FIG. 4. Results of chirped-pulse excitation measurements at 10 K with the pump pulse widths of (a) 2.4 ps and (b) 6.8 ps. Grey circles are cross-correlation signals between the chirped pump pulse and the unchirped 35-fs gate pulse, and grey solid lines represent Gaussian fitting to them. The pump pulse widths are estimated without performing the deconvolution of the gate pulse width. The color plots represent the transient evolution of the loss function spectra.

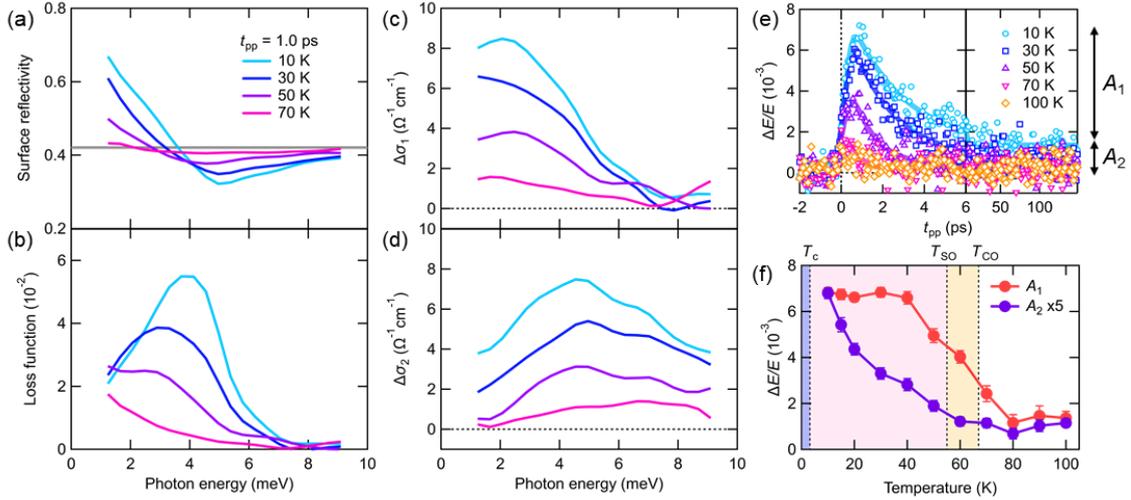

FIG. 5. Temperature dependence of the photo-induced reflectivity change taken at $t_{pp} = 1$ ps. (a) Surface reflection spectra, (b) loss function spectra $-\mathrm{Im}[1/\varepsilon(\omega)]$, and (c, d) real and imaginary parts of the conductivity change. Gray horizontal line in (a) represents the equilibrium reflectivity calculated from $n_{eq} = 4.7$. (e) Dynamics of THz electric field change for each temperature. Solid lines are fitting results by Eq. (1). (f) Temperature dependence of the fitting parameters $A_1$ and $A_2$ in Eq. (1).

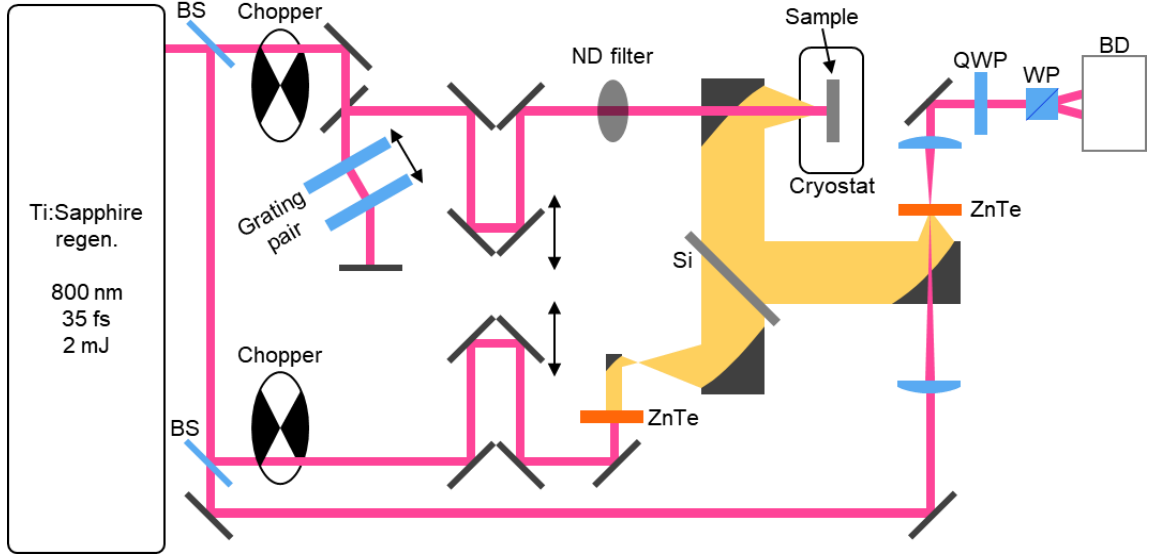

FIG. 6. Schematic image of the long-pulse excitation measurement. The optical pump pulse was down-chirped with a grating pair. BS: beam splitter, QWP: quarter wave plate, WP: Wollaston prism, BD: balanced detector.

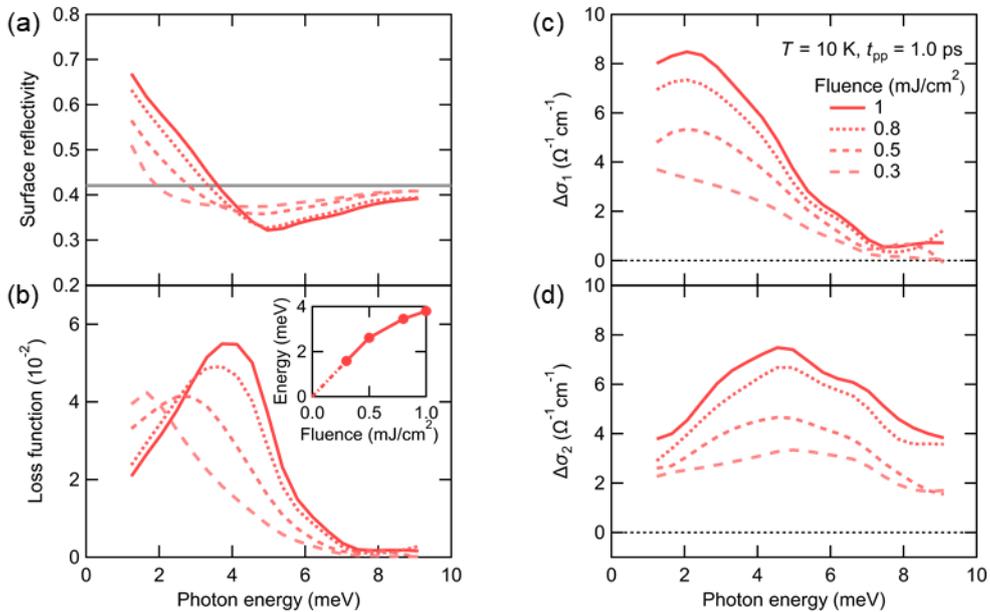

FIG. 7. (a) Surface reflection spectra, (b) loss function spectra $-\text{Im}[1/\varepsilon(\omega)]$, and (c, d) real and imaginary parts of the conductivity change at 10 K for several fluences. Gray line in (a) represents equilibrium reflectivity calculated from $n_{eq} = 4.7$. The inset of (b) shows the pump fluence dependence of the loss function peak energy at $t_{pp} = 1.0$ ps.